\documentclass[%
 reprint,
 amsmath,amssymb,
 aps,
]{revtex4-2}
  
\usepackage{graphicx}
\usepackage{dcolumn}
\usepackage{bm}

\usepackage[caption=false]{subfig}

\usepackage{url}
\usepackage{braket}
\usepackage{hyperref}
\hypersetup{
  colorlinks   = true,    
  urlcolor     = blue,    
  linkcolor    = blue,    
  citecolor    = blue,    
}

\usepackage{amsmath}

\begin{document}

\title{The effects of microscopic scattering on terahertz third harmonic generation in monolayer graphene} 

\author{Parvin Navaeipour}
\email{p.navaeipour@queensu.ca} 
\author{Marc M. Dignam}
\affiliation{Department of Physics, Engineering Physics \& Astronomy, Queen's University, Kingston, Ontario K7L 3N6, Canada}

\begin{abstract}
Due to its linear dispersion, monolayer graphene is expected to generate a third harmonic response at terahertz frequencies. There have been a variety of different models of this effect and recently it has been experimentally observed. However, there is still considerable uncertainty as to the role of scattering on harmonic generation in graphene. In this work, we model third-harmonic
generation in doped monolayer graphene at THz frequencies
by employing a nearest-neighbour tight-binding model in the
length gauge. We include optical phonon and neutral impurity
scattering at the microscopic level, and examine the effects of scattering on the third harmonic response. We also compare to the results of a
phenomenological semiclassical theory, using a field-dependent scattering time extracted from the simulation, and find a significantly lower third harmonic field than that found from the microscopic model. This demonstrates that third-harmonic
generation is much more sensitive to the nature of the
scattering than is the linear response. We also compare the results of our full simulation to recent experimental results and find qualitative agreement.

\end{abstract}

\maketitle

\section{Introduction} \label{sec_introduction}
The energy-momentum dispersion of graphene, a zero-bandgap two-dimensional semiconductor, is linear near the Dirac points; this leads
to a constant carrier speed. The fact that the speed of the carriers
is independent of crystal momentum results in an interesting nonlinear response of
graphene to terahertz (THz) radiation\cite{M1,Hed,Gu,Gla}. In particular, third harmonic generation of THz radiation should occur both due to the nonlinear relationship between the crystal momentum and the current density and due to the interaction between interband and intraband parts of the current densities\cite{naib2014, naib2015, Al_Naib_2015}. \par

There have been only a few reports of the experimental observation of third harmonic generation from monolayer
graphene at THz frequencies \cite{ Mik1,ishikawa}. The experimental work in 2014 by Bowlan \emph{et al.} \cite{bowlan} reported THG from graphene at THz frequencies for a $45$-layer doped sample. In a recent experiment by Hafez \emph{et al.} \cite{nature_hafez}, THz high-harmonic generation in a single layer of graphene at room temperature was reported. By using a low-noise, long-pulse, THz free-electron laser \cite{free_laser} and frequency filtering, they were able to generate the third, fifth, and seventh THz harmonics using THz fields of a few tens of $\rm{kV/cm}$. Although they did not employ a microscopic model of the response, their semi-empirical analysis of the results indicated that scattering was playing an important role in the generation of the harmonics. \par

There have been a number of theoretical approaches to modelling harmonic generation in graphene. Some theoretical investigations of the nonlinear response of graphene \cite{Mik1,ishikawa} have ignored scattering effects. At optical frequencies, this is a reasonable approximation. However, because typical scattering times in graphene are only a few tens of femtoseconds \cite{hwang, hwang2007, chen, hong, sarma}, this is not a valid approximation at THz frequencies, since the scattering times are generally shorter than the period of the field. Moreover, recent theoretical \cite{luke,nature_hafez} and experimental \cite{Hafez2014,Hafez2015,nature_hafez} results indicate that scattering can often enhance the nonlinearity in graphene. \par
Al-Naib \emph{et al.} \cite{naib2015} treated the nonlinear THz response of graphene in a density matrix formalism, using a semi-empirical model of carrier scattering and found that third-harmonic generation (THG) is strongly dependent on scattering. Given its strong role in the nonlinear response of graphene, it is therefore important to implement as accurate a scattering model as possible when modelling the nonlinear THz response of graphene. Although there have been a number of theoretical and experimental papers incorporating microscopic models of carrier scattering in graphene \cite{Knorr,luke}, to date, there has not been a systematic investigation into the relative roles of the intrinsic nonlinearity and different scattering processes on third harmonic generation in monolayer graphene at THz frequencies. \par 

There are four important scattering processes in graphene: neutral impurity scattering, charged impurity scattering,
phonon scattering, and carrier-carrier scattering. Hwang \emph{et al.} \cite{hwang} have shown that the carrier-carrier scattering time for an electron with an energy of $1~\rm{eV}$ above the Dirac points is about $100~\rm{fs}$ and is longer for energies close to the Dirac point. This is quite long relative to the other scattering processes in many samples, especially at room temperature. The scattering time of the charged impurity scattering is proportional to the Fermi energy and inversely to the density of impurities. Thus, one finds that charged impurity scattering can be very strong for low Fermi levels, while
neutral impurity scattering is strongest for high Fermi levels because the scattering time is inversely proportional to the carrier energy. \par 
E. Malic \emph{et al.} \cite{Knorr} employed a microscopic approach based on
a many-particle density matrix formalism to model carrier dynamics in \emph{optically}
excited graphene. By using this approach, it is possible to model time, momentum, and angle-resolved relaxation dynamics of non-equilibrium charge carriers during and
after excitation by an optical pulse. They considered the light-carrier
interaction as well as carrier-carrier and carrier-phonon scattering using their "graphene Bloch equations", which describe the time evolution of the carrier populations as
well as the interband microscopic polarization. \par 
In this work, we investigate the relative roles that the intrinsic nonlinearity and scattering play on the nonlinear response of doped monolayer graphene to THz fields. We employ a density
matrix formalism in the length gauge with microscopic scattering due to neutral impurities and optical phonons treated in the manner of E. Malic \textit{et al.} \cite{Knorr, rossi}. We show that indeed the scattering mechanisms play a strong role in third harmonic generation, but that their effect on the generated field is quite subtle, with an important interplay occurring between the elastic and inelastic scattering processes. In addition, we model the experiments of Hafez \emph{et al.} \cite{nature_hafez} and obtain qualitative agreement. \par 
The paper is organized as follows: In Sec. \ref{sec_SCtheory}, we present a semiclassical model of the nonlinear response of graphene with phenomenological scattering. In Sec. \ref{sec_microtheory}, we discuss our density matrix model with the inclusion of microscopic scattering from neutral impurities and optical phonons. In Sec. \ref{sec_Results}, we present and analyse the results of our density matrix simulations of the nonlinear response, compare it to the results of our semiclassical model and investigate the effects of different scattering mechanisms on the transmitted field at the fundamental and the generated third harmonic field as a function of THz field amplitude. In Sec. \ref{sec_Results_Expt} we compare our results for the generated third-harmonic electric field with the results obtained in the experiments of Hafez \emph{et al.} \cite{nature_hafez}. Finally, in Sec. \ref{sec_Conclusion}, we summarize our results. 
\section{Theory}\label{sec_theory}

In this section, we present our model of nonlinear carrier dynamics in graphene. We start with a simple semiclassical perturbative model of the nonlinear response of graphene to a harmonic THz field in the presence of phenomenological scattering. Then, in Sec. \ref{sec_microtheory}, we present our more complete non-perturbative microscopic theory that models the nonlinear response to THz pulses in the presence of scattering due to neutral impurities and optical phonons.

\subsection{Simple semiclassical model}\label{sec_SCtheory}

A very simple model of the intrinsic nonlinearity of graphene can be obtained using a semiclassical theory. The semiclassical equation for the statistical average of the electron wave vector is
\begin{equation}
\label{eq_Boltz}
\frac{d\mathbf{k}\left(  t\right)  }{dt}=\frac{e}{\hbar}\mathbf{E}_t\left(
t\right)-\frac{\mathbf{k}\left(  t\right)}{\tau},
\end{equation}
where $e$ is the charge on an electron, $\tau$ is a phenomenological scattering time, and $\mathbf{E}_t\left(  t\right)$ is the THz electric field at the graphene (\textit{i.e.} the transmitted field). In a nearest-neighbor tight-binding model, for energies less than about 800 meV, the conduction band energy for monolayer graphene is given by, $E_{c}\left( \mathbf{k}\right)  =\hbar v_F k$,
where $v_{F} = 1.0\times 10^6 m/s$ is the Fermi velocity and $k\equiv \vert\mathbf{k}\vert$. 
At zero temperature, for a Fermi energy of $E_F$, all states in a disk in $k$-space with radius $k_F = E_F /(\hbar v_F)$ are occupied. \par 
Let us now consider that the system is driven by a harmonic field such that the electric field at the graphene (the transmitted field) is given by $\mathbf{E}_t\left(  t\right)  =\hat{\mathbf{x}}{E}_{to}e^{-i\omega t} + c.c.$. At time $t=-\infty$, the Fermi disk is centered at the origin. Solving Eq. (\ref{eq_Boltz}), we see that the center of the disk at time $t$ is given by $\mathbf{k}_c\left( t\right)=k_{cx}\left( t\right)\hat{\mathbf{x}}$, where
\begin{equation}
    k_{cx}\left( t\right)=\frac{e}{\hbar}\frac{E_{to}e^{-i\omega t}}{\left (1/\tau-i\omega \right )  } + c.c.
\end{equation}
Now, assuming that the Fermi energy is high enough such that $2E_F \gg \hbar\omega$, there won't be any interband transitions and the x-component of the current density is given simply by
\begin{equation}
    \label{eq_current1}
J_x\left(  t\right)  =\frac{4e}{A}{\displaystyle\sum\limits_{\mathbf{k},occ}}
v_x\left(  \mathbf{k} \right),\\
\end{equation}
where $A$ is the area of the graphene, the factor of 4 accounts for the two spins and two valleys and
\begin{align}
    v_x\left(  \mathbf{k}\right)  &\equiv \frac{1}{\hbar}\frac{dE_{c}\left(  \mathbf{k}\right)}{dk_x}\\
    & =v_F \cos\left(\theta\right) \nonumber
\end{align} 
is the $x$-component of the carrier velocity, where $\theta$ is the angle of the $\mathbf{k}$ relative to the $x$-axis. Thus, converting the sum to an integral, the current density becomes
\begin{equation}
    \label{eq_current2}
J_x\left(  t\right)  =\frac{4ev_F}{\left(2\pi\right)^2}\int_0^{2\pi}d\theta \cos\left(\theta\right)
\int_0^{\infty}dk\,k\,\Theta \left(k_F-\vert\mathbf{k}-\mathbf{k}_c(t)\vert \right),
\end{equation}
where $\Theta\left(k\right)$ is the Heaviside function. If we assume that $\vert \mathbf{k}_c(t)\vert << k_F$, then we can expand the Heaviside function about $\mathbf{k}_c=0$, which gives 
\begin{align}
    \label{eq_Heaviside}
&\Theta\left(k_F-\vert\mathbf{k}-\mathbf{k}_c(t)\vert \right)  \approx \Theta\left(k_F-k \right)
+k_{cx}\left(  t\right)\cos\left(\theta\right)\delta \left(k_F-k \right)\nonumber\\
&\quad  +\frac{k_{cx}^2\left(  t\right)}{2!}\left[ -\frac{\sin ^2\left(\theta\right)}{k}\delta \left(k_F-k \right) +\cos ^2\left(\theta\right)\delta^{(1)} \left(k_F-k \right)\right]\nonumber\\
&\quad  -\frac{k_{cx}^3\left(  t\right)}{3!}\left\{ \frac{3\cos\left(\theta\right)\sin^2\left(\theta\right)}{k^2}\left[\delta \left(k_F-k \right) +k\delta^{(1)} \left(k_F-k \right)\right]\right.\nonumber\\
&\quad \left. -\cos^3\left(\theta\right)\delta^{(2)} \left(k_F-k \right)\right\}+O(k_{cx}^4),
\end{align}
where $\delta \left( k\right)$ is the Dirac delta function and $\delta^{(n)} \left( k\right)$ is its nth derivative. Using this in Eq. (\ref{eq_current2}) and performing the integral over $k$ and omitting the terms that are even in $k_{cx}\left(  t\right)$ (which will go to zero when we integrate over $\theta$), we obtain
\begin{align}
    \label{eq_current3}
J_x\left(  t\right)  &=\frac{4ev_Fk_F^2}{\left(2\pi\right)^2}\int_0^{2\pi}d\theta \cos\left(\theta\right)
\left[ \tilde{k}_{cx}\left(  t\right)\cos\left(\theta\right)\right.\nonumber\\  
&  -\frac{3\tilde{k}_{cx}^3\left(  t\right)}{3!}\cos\left(\theta\right)\sin^2\left(\theta\right)\nonumber\\
& \left. -\frac{15\tilde{k}_{cx}^5\left(  t\right)}{5!}\cos\left(\theta\right)\sin^4\left(\theta\right)+O(\tilde{k}_{cx}^7) \right],
\end{align}
where $\tilde{k}_{cx}\left(  t\right)\equiv k_{cx}\left(  t\right)/k_F$ and we have now also included the fifth-order term. In the limit that there is no scattering, it is straightforward to show that Eq. (\ref{eq_current3}) agrees with the result of Mikhailov and Ziegler\cite{Mik1}. The advantage of this expression, however, is that it explicitly shows which regions of $k$-space give the largest contributions to the linear and nonlinear response. In particular, we see that while the main contributions to the linear part of the current occur for $\theta$ near 0 and $\pi$, the main contribution to the third-order response occurs when $\theta$ is near odd multiples of $\pi /4$, and the main contribution to the fifth-order arises when $\theta=m\pi\pm \arccos(1/\sqrt{3})$. We shall return to this point when we discuss the results of our microscopic simulations of THG in Sec. \ref{sec_microtheory}.\par
Finally, we can perform the integrals over $\theta$ in Eq. (\ref{eq_current3}) and insert our expression for $k_{cx}\left( t\right)$, to obtain expressions for the linear, third order, and fifth-order conductivities.  The linear conductivity is given by
\begin{equation}
\label{eq_DrudeT0}
\sigma_{xx}^{(1)}(\omega) =\frac{e^2 E_F}{\pi\hbar^2 (1/\tau -i\omega)},   
\end{equation}
which agrees with the standard results at $T=0$ \cite{Mik3, drude1, drude2}. The diagonal element of the third order conductivity is 
\begin{equation}
\label{eq_sigma33w}
\sigma_{xxx}^{(3)}(3\omega) =\frac{-e^2 v_F^2}{8E_F^2 (1/\tau -i\omega)^2}\sigma_{xx}^{(1)}(\omega)
\end{equation}
for the third harmonic response and 
\begin{equation}
\label{eq_sigma3w}
\sigma_{xxx}^{(3)}(\omega) =\frac{-3e^2 v_F^2}{8E_F^2 \left(1/\tau^2+\omega^2\right)^2}\sigma_{xx}^{(1)}(\omega)
\end{equation}
for the response at the fundamental. In the limit that $\tau\rightarrow\infty$, these expressions for the third-order response agree with the results found by Mikhailov\cite{M1}. Finally, the diagonal element of the fifth-order conductivity is
\begin{equation}
\label{eq_sigma55w}
\sigma_{xxx}^{(5)}(5\omega) =\frac{-e^4 v_F^4}{64 E_F^4(1/\tau -i\omega)^4}\sigma_{xx}^{(1)}(\omega)
\end{equation}
for the fifth harmonic response, 
\begin{equation}
\label{eq_sigma53w}
\sigma_{xxx}^{(5)}(3\omega) =\frac{-5e^4 v_F^2}{64 E_F^4 \left(1/\tau^2+\omega^2\right)(1/\tau -i\omega)^2}\sigma_{xx}^{(1)}(\omega)
\end{equation} 
for the third harmonic response, and
\begin{equation}
\label{eq_sigma5w}
\sigma_{xxx}^{(5)}(\omega) =\frac{-10e^4 v_F^2}{64 E_F^4 \left(1/\tau^2+\omega^2\right)^2}\sigma_{xx}^{(1)}(\omega)
\end{equation}
for the response at the fundamental.\par 
Note that up to fifth order, the current density at the fundamental will be given by $J_x\left(\omega;  t\right) =Re\left\{ 2\sigma_{NL}\left(\omega\right)E_{t0}e^{-i\omega t}\right\}$, where

\begin{equation}
\label{eq_SigNL}
    \sigma_{NL}\left(\omega\right)\equiv\sigma_{xx}^{(1)}(\omega)\left[
    1-3\left\lvert \frac{E_{tm}}{E_S}\right\rvert^2-10\left\lvert \frac{E_{tm}}{E_S}\right\rvert^4\right] 
\end{equation}
is the effective nonlinear conductivity, where
\begin{equation}
\label{eq_ES}
    E_S\equiv \frac{E_F\sqrt{32\left(1/\tau^2+\omega^2\right)}}{\lvert e\rvert v_F}
\end{equation}
is a saturation field, and $E_{tm}\equiv 2 E_{to}$ is the amplitude of the transmitted field. Similarly, the current density at the third harmonic will be given by $J_x\left(3\omega;  t\right) =Re\left\{ 2\sigma_{NL}\left(3\omega\right)E_{t0}^3e^{-i3\omega t}\right\}$, where
\begin{equation}
\label{eq_Sig3wNL}
    \sigma_{NL}\left(3\omega\right)\equiv \sigma_{xxx}^{(3)}(3\omega)\left[
    1+5\left\lvert \frac{E_{tm}}{E_S}\right\rvert^2\right]. 
\end{equation}

Note that for the conductivity at the fundamental frequency, the nonlinear components add out of phase with the linear component and so decrease the current density at the fundamental, which results in an increase the transmission as the field amplitude increases. As we will see later, for the doping densities, THz frequency, and field amplitudes we will consider in this work, if there is no scattering, then the intrinsic nonlinear reduction in the transmission at the fundamental frequency as predicted by this model can be appreciable for large field amplitudes. However, for small scattering times on the order of a few tens of femtoseconds, it has a very modest effect. For example for a Fermi energy of 354 meV, a frequency of 1.0 THz, and a THz field amplitude of 30 kV/cm, the nonlinearity only lowers the conductivity by 1.5\% for $\tau=50$ fs. In contrast, for a scattering time of 200 fs, it lowers the conductivity by 10\%, which is quite significant. This means that if the scattering times are say 100 fs or longer, then one cannot neglect the intrinsic nonlinearity when calculating the dependence of the transmitted power as a function of the incident pulse amplitude. \par  
In the next section, we present the results of a full dynamic model, including a microscopic treatment of scattering. As we shall see, in this model the transmission is predicted to change considerably with field amplitude due in large part to the energy dependence of the scattering times as well as scattering-induced carrier redistribution.

\subsection{Full microscopic theory}\label{sec_microtheory}

As in the previous section, we model n-doped graphene where THz-field-induced interband transitions can be neglected. We consider samples where short-range neutral impurity scattering dominates over charged-impurity scattering and scattering from acoustic phonons \cite{fang,hadi}. However, in addition to neutral impurity scattering, we include scattering from optical phonons, as this is the dominant inelastic scattering mechanism for electrons that are exited to high energies. For simplicity, we neglect carrier-carrier scattering, which is a reasonable approximation when the doping is not too heavy. Omitting field-driven interband transitions, the dynamical equation for the conduction band carrier density matrix is \cite{naib2014, naib2015}
\begin{align}
\label{eq_rhoccdynam}
\dfrac{d \rho_{cc}(\mathbf{k})}{dt} = -\dfrac{e\mathbf{E}_t(t)}{\hbar} \cdot \nabla_{\mathbf{k}} \rho_{cc} ( \mathbf{k}) - \Big( \dfrac{d \rho_{cc}(\mathbf{k})}{dt} \Big)_{scatt},
\end{align} 
where $\Big( \dfrac{d \rho_{cc}(\mathbf{k})}{dt} \Big)_{scatt}$ represents the time variation of carrier density due to the scattering and $\mathbf{E}_t(t)$ is the THz field at the graphene. \par

We again take the electron dispersion relation to be $E_c( \mathbf{k}) = \hbar v_F k$, where $v_F$ is $ ( = 1.0 \times 10^{6}~\rm{m/s})$ the Fermi velocity and we take the origin to be at the Dirac point. We include the electron-phonon, and electron-neutral-impurity scattering in the dynamic equations in the Born-Markov approximation. Thus, the electron scattering term is given by \cite{Knorr,luke}
\begin{align}
\Big( \dfrac{d \rho_{cc}(\mathbf{k})}{dt} \Big)_{scatt} = - \Gamma_c^{out} ( \mathbf{k}) \rho_{cc}(\mathbf{k}) + \Gamma_c^{in}( \mathbf{k}) [ 1 - \rho_{cc}( \mathbf{k})]. 
\end{align}  
In this equation, $ \Gamma_c^{out} ( \mathbf{k})$ is the scattering-out rate, and $ \Gamma_c^{in} ( \mathbf{k})$ is the scattering-in rate, which is given by 
\begin{align}
\label{eq_Gammain}
\Gamma_c^{in} ( \mathbf{k}) = \dfrac{2 \pi }{\hbar} \sum_{\mathbf{q}} \Big \lbrace  \sum_{j}\vert g_{\mathbf{q}j}^{\mathbf{k}cc} \vert ^2 \rho_{cc} ( \mathbf{k} + \mathbf{q}) [ n_j( \mathbf{q}) +1] \nonumber\\ 
\delta[\varepsilon_c( \mathbf{k} + \mathbf{q}) - \varepsilon_c( \mathbf{k}) - \hbar \omega_j( \mathbf{q}) ]  \nonumber\\
+ \sum_{j}\vert g_{\mathbf{q}j}^{(\mathbf{k}-\mathbf{q})cc} \vert ^2 \rho_{cc} ( \mathbf{k} - \mathbf{q})  n_j ( \mathbf{q}) \nonumber \\ 
\delta[ \varepsilon_c( \mathbf{k} - \mathbf{q}) - \varepsilon_c( \mathbf{k}) + \hbar \omega_j( \mathbf{q}) ] \nonumber \\
+\vert h_{\mathbf{q}}^{\mathbf{k}c} \vert ^2 \rho_{cc}( \mathbf{q})  \delta[ \varepsilon_c( \mathbf{q}) - \varepsilon_c ( \mathbf{k}) ] \Big \rbrace,  
\end{align}
where $j$ labels the phonon branch, $n_j(\mathbf{q})$ is the phonon occupation number, and $\hbar\omega_j(\mathbf{q})$ is the phonon frequency. The first and second terms in Eq. (\ref{eq_Gammain}) respectively represent scattering due to optical phonon emission and absorption. In the case of phonon absorption, the electrons are scattered to higher energy states in the conduction band due to the absorption of an optical phonon and in the second case, the electrons that have enough energy to stimulate the emission of a phonon are scattered to a lower energy state. The third term in Eq. (\ref{eq_Gammain}) represents neutral impurity scattering. The scattering-out rate can be obtained from $\Gamma_c^{in}(\mathbf{k})$ by replacing $\rho_{cc}(\mathbf{q})$ with $(1-\rho_{cc}(\mathbf{q}))$ and interchanging $n_j$ and $(n_j + 1)$. \par
In Eq. (\ref{eq_Gammain}), we include scattering due to longitudinal and transverse optical phonon with wave vectors close to the $\Gamma$-point as well as transverse optical phonons near the $K$-point. The squares of the coupling constants to these phonons are given respectively by
\begin{align}
\vert g_{\mathbf{q}\Gamma-LO}^{\mathbf{k}cc}\vert^2 =& \dfrac{1}{N} g_{\Gamma}^2 [ 1 - \cos( \theta_{\mathbf{q},\mathbf{k}} + \theta_{\mathbf{q},\mathbf{k}+\mathbf{q}})],\\
\vert g_{\mathbf{q}\Gamma-TO}^{\mathbf{k}cc}\vert^2 =& \dfrac{1}{N} g_{\Gamma}^2 [ 1 + \cos( \theta_{\mathbf{q},\mathbf{k}} + \theta_{\mathbf{q},\mathbf{k}+\mathbf{q}})], \\
\vert g_{\mathbf{q}K}^{\mathbf{k}cc} \vert^2 =& \dfrac{1}{N} g_K^2 [ 1 - \cos(\theta_{\mathbf{k},\mathbf{k}+\mathbf{q}})],
\end{align}
where $N$ is the number of unit cells, $\theta_{\mathbf{k},\mathbf{q}}$ is the angle between $ \mathbf{k}$ and $ \mathbf{q}$, and $g_{\Gamma}^2$ and $ g_K^2$ are the squares of the amplitudes of the coupling constants. There is still uncertainty in the experimental and theoretical literature \cite{impact, Robertson2004, Robertson2008} as to what one should use for the values for these coupling constants. However, we use the values $g_{\Gamma}^2 = 0.0405~\rm{eV}^2$, and $ g_K^2 = 0.0994~\rm{eV}^2$ given in Ref. \cite{Knorr} to be consistent with the calculations of Helt \textit{et al.} \cite{luke}. We take the optical phonons to be dispersionless near the symmetry points, so that for the phonons near the $ \Gamma$- point, $\hbar \omega_{\rm{\Gamma-LO}}( \mathbf{q}) \approx \hbar \omega_{\rm{\Gamma-TO}}( \mathbf{q}) \approx \hbar \omega_{\rm{\Gamma}} \approx 196~\rm{meV}$, while for optical phonons near the ${K}$-point, $\hbar \omega_K ( \mathbf{q}) \approx \hbar \omega_K = 160~\rm{meV}$ \cite{fang}. As the THz pulses are relatively short, we do not calculate the phonon dynamics, but rather assume that they are in thermal equilibrium, such that $ n_{j} = [\exp[\beta\hbar \omega_{j}]-1]^{-1}$, where $\beta\equiv [k_B\rm{T}]^{-1}$, where $\rm{T}$ is the lattice temperature and $k_B$ is the Boltzmann constant.  \par

In the final term in Eg. (\ref{eq_Gammain}), 
the square of the carrier-neutral impurity coupling element is given by \cite{hwang}
\begin{equation}
\vert h_{\mathbf{q}}^{\mathbf{k}c} \vert^2 = \dfrac{n_{imp} v_0^2}{A} [ 1 + \cos(\theta_{\mathbf{k},\mathbf{q}})],
\end{equation}
where $n_{imp}$ is the neutral impurity density and $v_0$ is a constant interaction strength as appropriate for short-range point defect scatters. In what follows, we take $v_0 = 1$ keV ${\r{A}}^2$, as given in Ref. \cite{sarma}. \par
We take the incident THz field to be a single-cycle sinusoidal pulse with a Gaussian envelope given by 
\begin{equation}
\widehat{\mathbf{E}}_i(t) = \dfrac{E_0}{N_E} \exp \Big\lbrace {- \dfrac{4 \rm{ln}(2) (\emph{t} -\emph{t}_0)^2}{T_{\rm{FWHM}}^2}} \Big\rbrace \sin[ 2 \pi f_0 ( t - t_0) ] ~\widehat{\mathbf{e}}_x,
\end{equation}
where $E_0$ is the peak field amplitude, $T_{\rm{FWHM}}$ is the full width at half maximum of the pulse, $t_0$ is the time offset, and $f_0$ is the pulse carrier wave frequency. The constant $N_E$ depends on the pulse duration and is chosen such that $E_0$ is the peak THz field amplitude. \par 
Because we consider n-doped graphene in this work, the only current is the intraband current in the conduction band, which is given by \cite{riely} 
\begin{equation}
\mathbf{J} = 4e v_F \sum_{\mathbf{k}} \rho_{cc}(\mathbf{k}) ~\widehat{\mathbf{k}}.
\end{equation} 
The electric field transmitted through the monolayer graphene sheet on a substrate with refractive index $n$ is given by 
\begin{equation}
\label{eq_Etrans}
\mathbf{E}_t(t) = \dfrac{2\mathbf{E}_i(t) - Z_0 \mathbf{J}[\mathbf{E}_t(t)]}{1 + n},
\end{equation} 
where $\mathbf{J}[\mathbf{E}_t(t)]$ is the total current density calculated using
the transmitted field as the driving field and $Z_0$ is the impedance of free space. Thus, solving the dynamic equations for the density matrix requires that the field at the graphene is calculated self-consistently, as described by Al-Naib \textit{et al.} \cite{naib2014}. \par
To solve Eq. (\ref{eq_rhoccdynam}), we discretize $k$ on a hexagonal grid \cite{luke} and solve the coupled dynamic equations using the Runge-Kutta algorithm. Because the Dirac points at $K$ and $K'$ are identical, we only need to perform our calculations about the $K$ point. We find that a grid size of $601~\times~601$ is required for convergence of our results for the Fermi energies and THz field pulse amplitudes considered. The most computationally intensive part of the calculation is the evaluation of the scattering terms. The numerical approach used to evaluate these is described by Helt and Dignam \cite{luke}. To ensure that no carriers are driven outside of the simulation grid due to the scattering processes or applied field, we set the grid edge, Max$\{\vert \mathbf{k} \vert\} $, to be $1.5$ times the maximum displacement of the edge of the electron disc when driven by the strongest incident field $\mathbf{E}_i(t)$ (in the absence of scattering) \cite{luke}. We take the time duration of the simulation to be long enough to obtain convergence and sufficient frequency resolution of the transmitted field. \par
Helt and Dignam \cite{luke} studied the effect of scattering on the nonlinear transmission close to the fundamental frequency as a function of the THz field amplitude. In this work, we are primarily interested in the effect of scattering on \emph{third harmonic generation}. In the following section, we investigate the nonlinear response of graphene with microscopic scattering. To better understand the effects of the different scattering mechanisms, we present results for the frequency spectrum of the transmitted electric field when there is no scattering, only phonon scattering, only impurity scattering, both phonon and impurity scattering, and when we use an empirical scattering model with an energy-independent scattering time. We also compare our simulation results with the results of the simple semiclassical model of Sec. \ref{sec_SCtheory}. In Sec. \ref{sec_Results_Expt}, we model the experiments of Hafez \emph{et al.}. \cite{nature_hafez} and discuss the generated THz fields and the carrier distributions for that system. 
\section{Results}\label{sec_Results}
We now present the results of our simulations of the nonlinear THz response for n-doped monolayer graphene. In this section, we take the central frequency of the incident THz pulse to be $f_0=1.0~\rm{THz}$ and take the pulse duration to be $1.0~\rm{ps}$. We take the graphene to be suspended (such that $n = 1$), the temperature to be $300~\rm{K}$, the chemical potential to be $\mu_{c}=354~\rm{meV}$ and the neutral impurity density to be $n_{imp} = 3 \times 10^{10}~\rm{cm}^{-2}$. We have chosen these values because they correspond approximately to the field frequencies, carrier densities and low-field scattering times in a number of recent experiments on the nonlinear response of graphene \cite{Hafez2014,Hafez2015,hadi}. In addition, they are the parameters use in a recent paper of ours on the effects of neutral impurity and optical phonon scattering on the nonlinear transmission of graphene \cite{luke}. \par
To better understand the effect of the different scattering mechanisms, it will be useful to compare our results to that using a semi-empirical model of scattering involving only one energy-independent scattering time, $\tau$. The model that we shall use is one where the carriers relax back to thermal equilibrium over this time.  Thus we use 
\begin{equation}
   \Big( \dfrac{d \rho_{cc}(\mathbf{k})}{dt} \Big)_{scatt}^{empirical} =-\frac{[\rho_{cc}(\textbf{k}) -f_{FD}(\textbf{k})]}{\tau},
\end{equation}
where $f_{FD}$ is the Fermi-Dirac distribution for conduction band electrons. Solving the dynamic equations with this semi-empirical scattering term to first order in the THz field gives the standard temperature-dependent Drude conductivity, \cite{Mik3, drude1, drude2} 
\begin{equation}
\label{eq_Drude}
\sigma(\omega) =\frac{2e^2\ln{\left[2\cosh({\frac{\beta\mu_c}{2}}) \right]}}{\pi\beta \hbar^2 (1/\tau -i\omega)}.   
\end{equation}
At room temperature, for our Fermi energy, this expression agrees with Eq. (\ref{eq_DrudeT0}) derived in Sec. \ref{sec_SCtheory} to better than one part in $10^7$. \par 
At room temperature and for low field amplitudes, neutral impurity scattering is the dominant mechanism. Although the scattering rates are $k$-dependent, we can estimate the effective low-field scattering rate by evaluating the scattering out rate for carriers at the energy of the chemical potential. From Eq. (\ref{eq_Gammain}) and including only the neutral impurity scattering term, with $ \vert \mathbf{k} \vert = \mu_c / ( \hbar v_F) $ and $ \rho_{cc} = 1/2$, we find the low-field scattering time to be given by  
\begin{align}
\dfrac{1}{\tau} =&  \dfrac{n_{imp} v_0^2}{2 \pi \hbar^2 v_F} \int_0^{2 \pi}  d \theta_q \dfrac{\mu_c}{\hbar v_F} \dfrac{1}{2} \nonumber\\
=& \dfrac{n_{imp} v_0^2 \mu_c }{2 \hbar^3 v_F^2}. 
\end{align}
This is the scattering time that we shall use in our semi-empirical model. It should give good agreement for the linear response at low-field amplitudes. For our choice of chemical potential and impurity density, we obtain $\tau =  52~\rm{fs}$.  \par
In this section, we examine the effect of the different scattering mechanisms on the nonlinear transmission at the fundamental frequency and on the generated third harmonic. Thus, we simulate the transmitted fields as a function of time for different scattering mechanisms and also for different input field strengths and Fourier transform the results to obtain the incident and transmitted fields, ($  \mathbf{E}_i(\omega) $ and $  \mathbf{E}_t(\omega)  $) as a function of frequency. 
To make clear the role of each scattering process, we perform separate simulations under the following conditions: 1) no scattering; 2) only impurity scattering; 3) only optical phonon scattering; and 4) both phonon and impurity scattering. We also compare these results with those found using our semi-empirical model of scattering. \par
\begin{figure}
  \centering
  \includegraphics[scale=0.35]{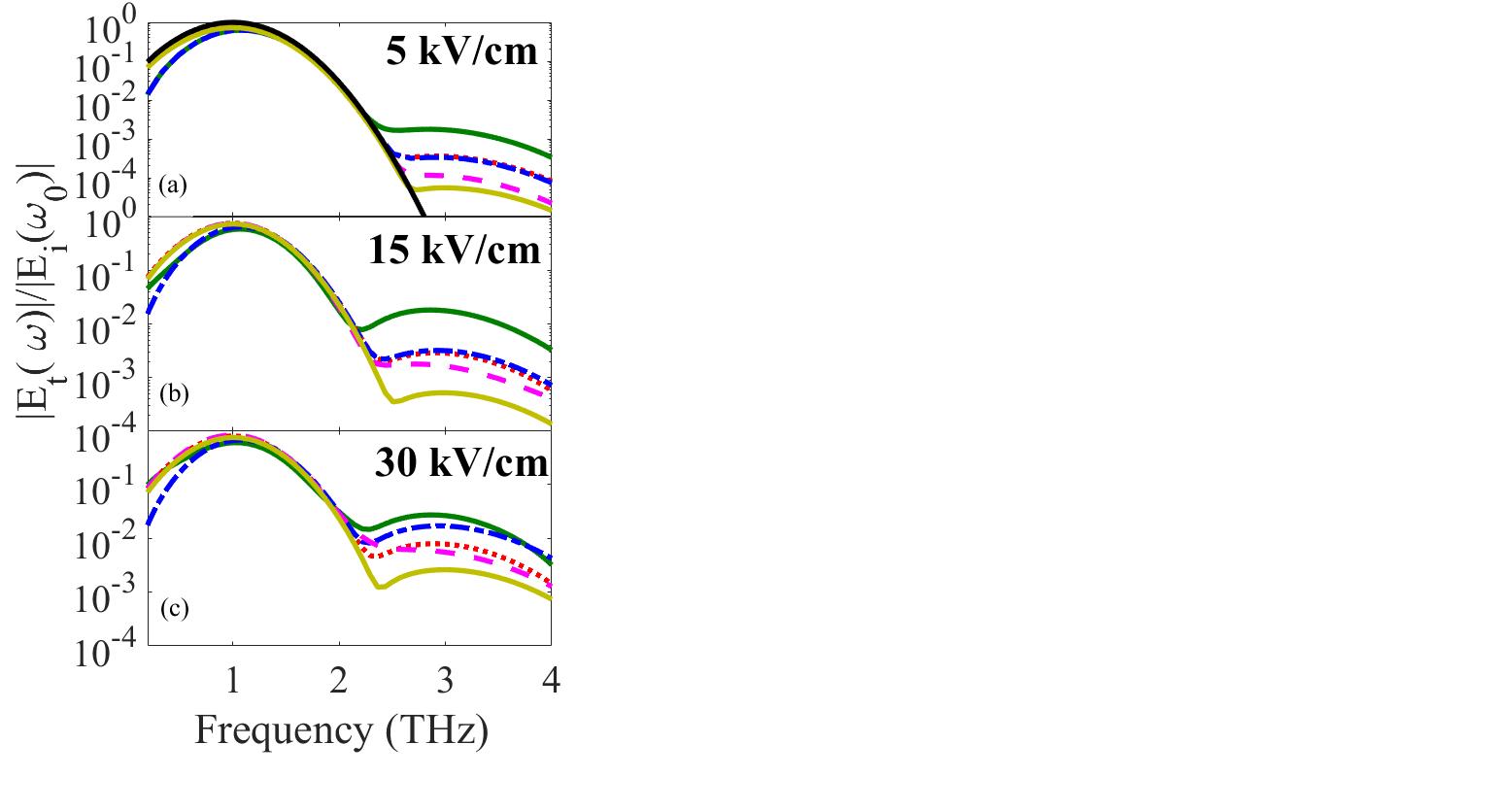} 
  \caption[Normalized transmitted field for different scattering mechanisms]{Normalized transmitted field for different scattering mechanisms when the incident field peak amplitude is (a) $5~\rm{kV/cm}$, (b) $15~\rm{kV/cm}$, and (c) $30~\rm{kV/cm}$. The black solid curve in (a) shows the normalized incident field with central frequency of $1~\rm{THz}$. The black curve in (a) is the normalized incident field. The other curves represent the normalized transmitted field with no scattering (dashed-dotted blue curve), only phonon scattering (solid dark green), only impurity scattering (dashed pink curve), impurity and phonon scattering (dotted red curve), and the results of the semi-empirical model (solid light green curve).}
  \label{fig:figure64}
\end{figure}

In Fig. \ref{fig:figure64}, we plot the transmitted field normalized to the peak of the incident field at $\omega=\omega_0=2\pi f_0$, ${\vert E_t( \omega) \vert}/{\vert E_i(\omega_0) \vert }$, as a function of frequency for the different scattering mechanisms (as indicated in the caption) for incident field amplitudes of (a) $5~\rm{kV/cm}$, (b) $15~\rm{kV/cm}$, and (c) $30~\rm{kV/cm}$ respectively.   
For reference, the solid black curve in Fig. \ref{fig:figure64}(a) gives the normalized incident field with central frequency of $1~\rm{THz}$. We see that the incident field only has a peak at $f = f_0 = 1~\rm{THz}$. However, the transmitted field for all the scattering mechanisms also has a peak near $3~\rm{THz}$. This peak represents the third-harmonic field, in which we are primarily interested. As can be seen, the generated third-harmonic field amplitude is strongly dependent on the scattering mechanism, while the transmitted field at $f_0$ has a weaker dependence, in large part because it is dominated by the transmission of the incident field. In Fig. \ref{fig:figure64}, the dashed blue curve gives the result when there is no scattering, which is obtained by setting $\Gamma_c^{in} ( \mathbf{k}) = \Gamma_c^{out} ( \mathbf{k}) = 0$. The solid green curve represents the normalized transmitted field including only phonon scattering, \emph{i.e.}, we omit the last term in Eq. (\ref{eq_Gammain}). The dashed purple curve gives the normalized transmitted field when only neutral impurity scattering is included. The dotted red curve shows the normalized transmitted field when both neutral impurity scattering and optical phonon scattering are included. Finally, the light blue curve gives the result obtained using the semi-empirical model. For the semi-empirical model we set the phenomenological scattering to be $52~\rm{fs}$ so that it agrees with the time obtained from neutral impurity scattering for low field amplitudes. \par
\subsection{Transmitted field at the fundamental frequency}\label{sec_Results_Omega}
Before examining in detail the effects of scattering on the \textit{third harmonic}, we first examine the nonlinear transmission of the graphene at the \textit{fundamental frequency} $\omega_0$. This was examined by Helt and Dignam \cite{luke} for a few different field amplitudes. Our goal here, however is to compare the results of our simulations with the results from our semiclassical model and to thereby extract effective scattering times as a function of field amplitude for a range of amplitudes.\par
\begin{figure}
  \centering
  \includegraphics[scale = 0.25]{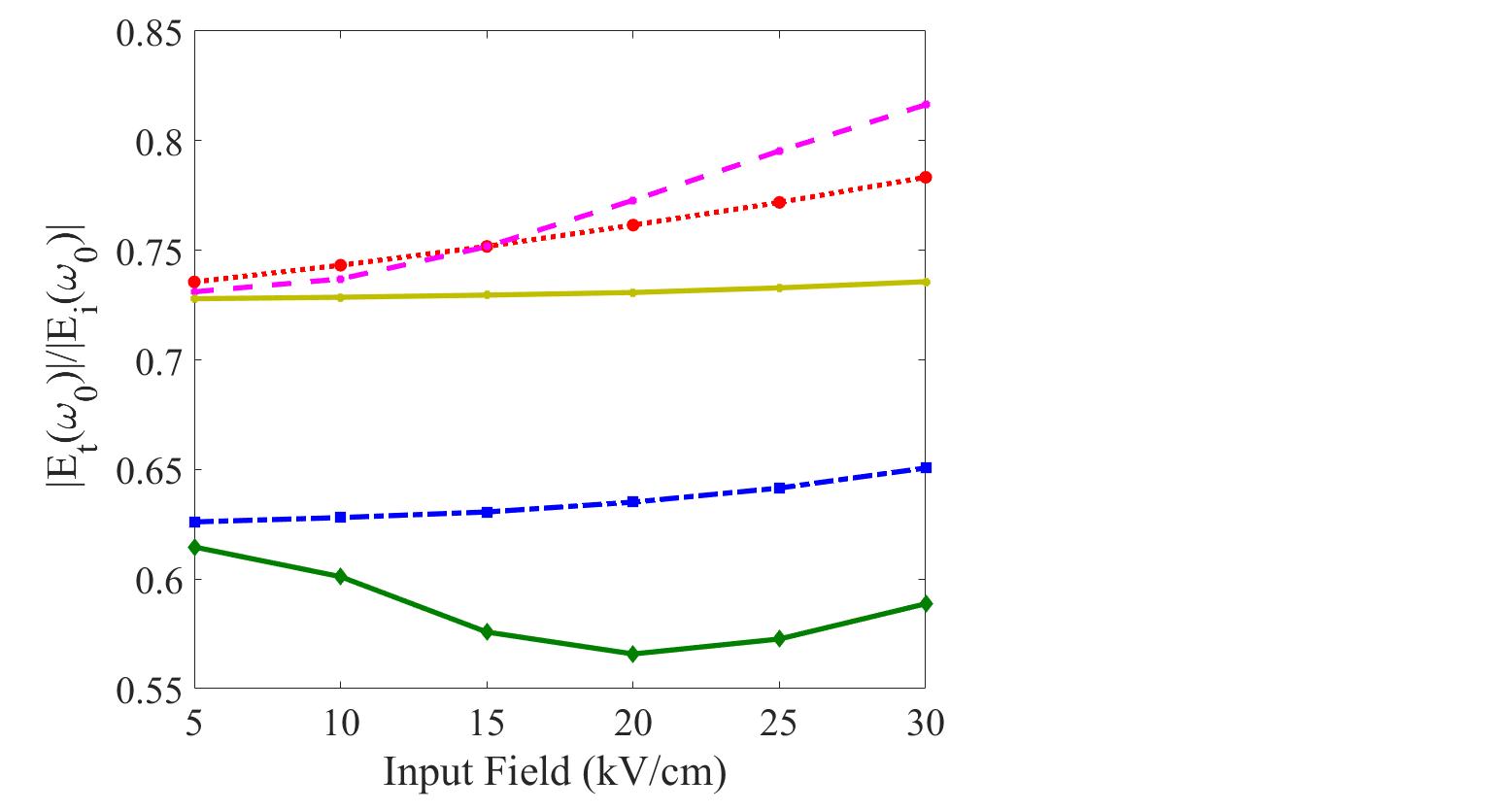} 
  \caption[Ratio of the transmitted field at $\omega_0$ to the peak value]{ Ratio of the transmitted field at $\omega_0$ to the peak value of the incident field as a function of input field for different scattering treatments; only phonon scattering (solid dark green), no scattering (dashed-dotted blue curve), impurity and phonon scattering (dotted red curve), impurity scattering (dashed pink curve), and semi-empirical model (solid light green curve).}
  \label{fig:pa}
\end{figure}

 
In Fig. \ref{fig:pa} we plot the ratio of the transmitted field at $\omega_0$ to the incident field at $\omega_0$ as a function of the amplitude of the input field for different scattering mechanisms. 
We note first that, as expected, this ratio depends both on the input field and the scattering mechanisms. This is because the transmission depends on the conductivity of the graphene, which in turn depends on the scattering and field amplitude.\par

\begin{figure}
  \centering
  \includegraphics[scale = 0.24]{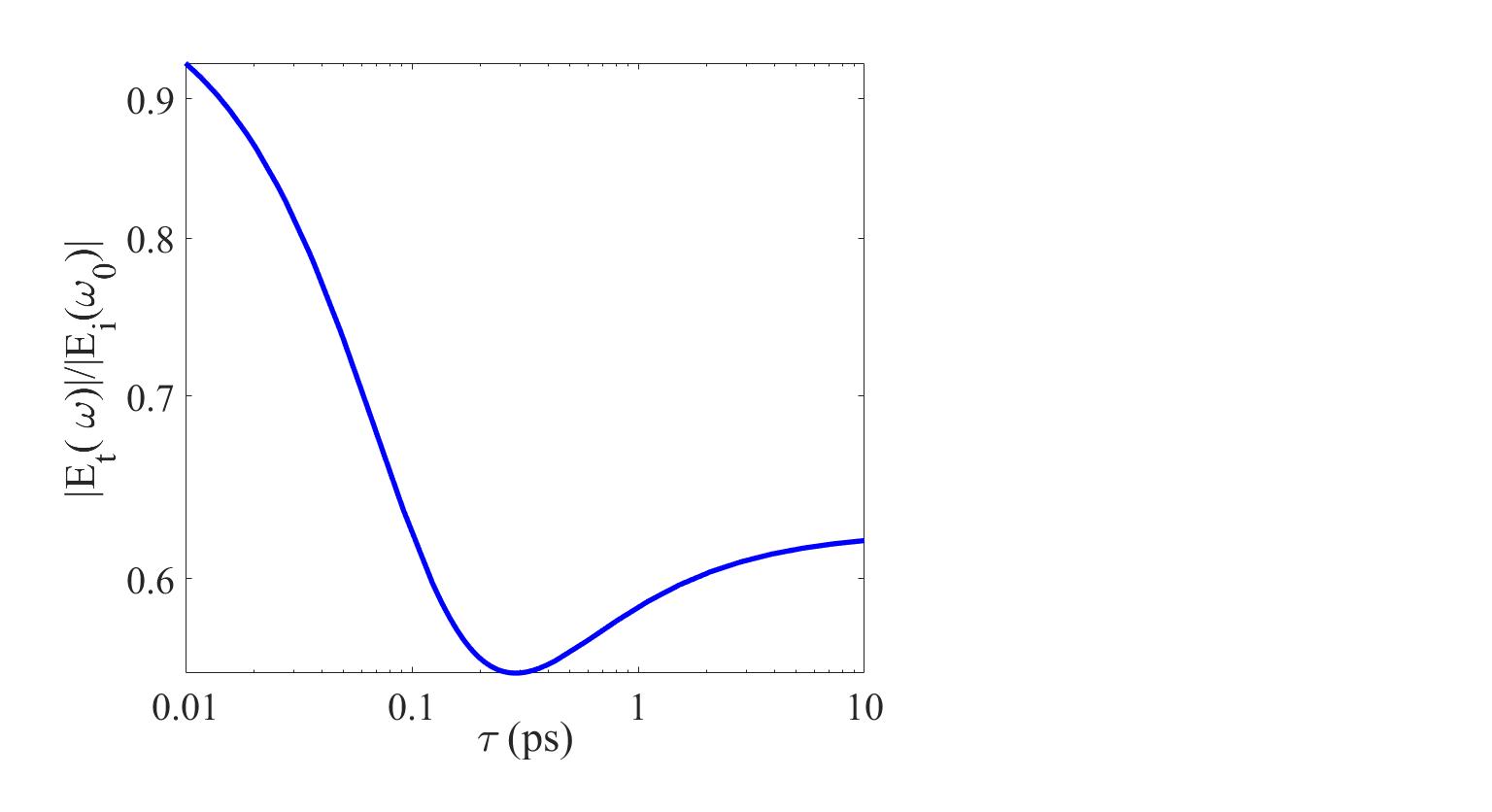} 
  \caption[Normalized transmitted field using the Drude model.]{Normalized transmitted field as a function of scattering time using the Drude model.}
  \label{fig:Transmission Analytic}
\end{figure} 
We first consider the effect of the different scattering mechanisms on the low-field ($E_0=5~\rm{kV/cm}$) transmission. To extract the low-field scattering times, we use the Drude model, but with a scattering time that depends on the particular scattering mechanism. From Eq. (\ref{eq_Etrans}), with $J_x(\omega)=\sigma (\omega) E_t (\omega)$, with $\sigma (\omega)$ given by the Drude model of Eq. (\ref{eq_Drude}), we can solve for the transmitted field to get the standard thin film result,
\begin{equation}
\label{eq_Etsimple}
    E_t(\omega)=\frac{E_i(\omega)}{1+\frac{1}{2}Z_o\sigma(\omega)}.
\end{equation}
Using this equation and the Drude model, we calculate the transmission, and in Fig. \ref{fig:Transmission Analytic}, we plot $\vert  E_t(\omega_0)/ E_i(\omega_0)\vert$ as a function of $\tau$ for the chemical potential and temperature that was used in the simulations. We can use this model to extract the effective low-field scattering time for each of the different scattering mechanisms. \par 

When there is only phonon scattering, we expect to obtain very similar transmission to the no-scattering case at low field amplitudes, since the phonon emission and absorption are negligible. This is due to the fact that all of the electron states below the occupied states are full, so scattering due to phonon emission is forbidden. In addition, because the phonon populations at room temperature are very small ($n_{\Gamma} = 0.00051$, $n_{\rm{K}} = 0.0021$), scattering via phonon absorption is also very weak. Thus, as expected, the transmission with only optical phonons is similar to what we obtained with no scattering. From Fig. \ref{fig:Transmission Analytic}, we extract an effective low-field scattering time due to optical phonons alone of about $\tau =4.4$ ps. \par 

When we include only neutral impurity scattering, the transmission is considerably increased, due to the strong damping of the carrier response. From Fig. \ref{fig:Transmission Analytic}, we extract an effective low-field scattering time due to neutral impurities alone of about $\tau =52$ fs, as expected. \par 

When both phonon and impurity scattering are included, the transmission essentially is the same as with only neutral impurity scattering, since the effect of phonon scattering is small in the case both neutral impurity and phonon scattering are present. \par 

The results using the semi-empirical model are also almost identical to the results found with neutral impurity scattering because we have chosen a scattering time based on low-field neutral impurity scattering. \par

Now let us examine what happens to the transmission at the fundamental frequency when we increase the input field amplitude. There are two effects at play here: the intrinsic nonlinearity and the energy-dependent scattering rates. First, as we discussed in Sec. \ref{sec_SCtheory}, due to the linear dispersion of the electron bands, the electron velocity is not proportional to the crystal momentum and so there will be "clipping" of the THz-induced current at high field amplitudes \cite{Al_Naib_2015, ishikawa, bowlan, M1, Mik1}. Second, as we shall see, the scattering mechanisms can both dampen the amplitude of carrier oscillation, while introducing a nonlinearity due to the energy dependence of the scattering rate. \par
\begin{figure}
    \centering
    \includegraphics[scale = 0.25]{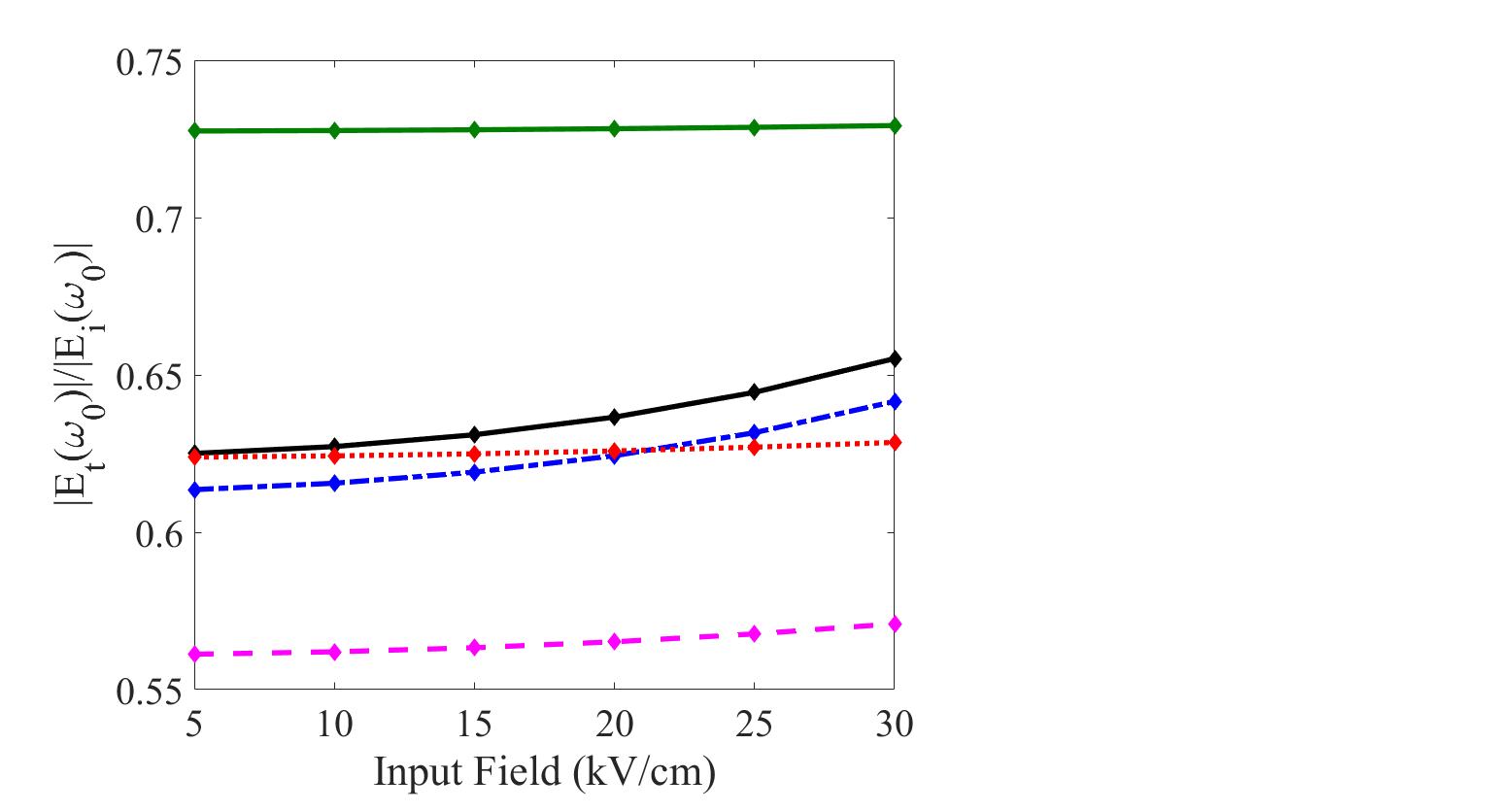}
    \caption[Ratio of the transmitted field]{Ratio of the transmitted field at $\omega_0$ to the peak value of the incident field as a function of input field as calculated using the simple nonlinear, semiclassical model for scattering times of $\tau=\infty$ (solid black curve), $\tau=4$ ps (dashed-dotted blue curve), $\tau=200$ fs (dashed pink curve), $\tau=100$ fs (dotted red curve), and $\tau=52$ fs (solid green curve).}
    \label{fig:Semiclassical}
\end{figure}

 
To aid in the discussion of the nonlinear effects due to scattering, in Fig. \ref{fig:Semiclassical}, we plot (for five different scattering times), the transmission calculated using Eq. (\ref{eq_Etsimple}) with $\sigma\left(\omega\right)$ replaced by the nonlinear semiclassical conductivity given by Eq. (\ref{eq_SigNL}). Note that because the nonlinear conductivity depends on the transmitted field, we need to iterate this to convergence. Note also that this result is for monochromatic fields and so we don't expect perfect agreement with the results of the full simulations, even when the scattering mechanism is the same.\par 

When there is no scattering, the semiclassical saturation field (see Eq. (\ref{eq_ES})) is $E_S\approx$ 126 kV/cm and the intrinsic nonlinearity is expected to have a significant effect at the higher fields. The simple semiclassical model predicts that the transmission at the highest field of 30 kV/cm will be about 4.8\% higher than at 5 kV/cm. This is in good qualitative agreement with the results from the simulation, which give a 3.9\% increase (see Fig. \ref{fig:pa}). \par 

For the semi-empirical model, where the scattering time is 52 fs, the semiclassical saturation field is 405 kV/cm and so the nonlinear effect is much smaller. The semiclassical model gives a 0.23\% increase in the transmitted field for an input field amplitude of 30 kV/cm relative to that at 5 kV/cm, which is in qualitative agreement with the full simulation, which gives an increase of 1.1\%. The agreement in this case is not as good because in the semiclassical model, we do not account for the scattering-in to low energy states. The change in the transmission with field is very small because scattering greatly limits how far in $k$-space the carriers are driven and we have assumed that the scattering time is energy-independent. \par 

The results for the change in transmission with field when there is microscopic scattering is not described at all well by the simple semiclassical model with a \textit{field-independent} scattering time. For example, when there is only impurity scattering, when we compare the highest input field to the lowest, the transmission increases by 12\% in the simulation results, but if we keep the scattering time at the low field value, the semiclassical model predicts an increase of only 0.23\%. Even more striking is the very different behaviour of the transmission simulation results when there is optical phonon scattering from any results found using a constant scattering time. In the simulation, the transmission first decreases and then increases as the field amplitude is increased, while in the semiclassical model, there is always an increase in the transmission as the input field amplitude is increased. \par
\begin{figure}
  \centering
  \includegraphics[scale = 0.23]{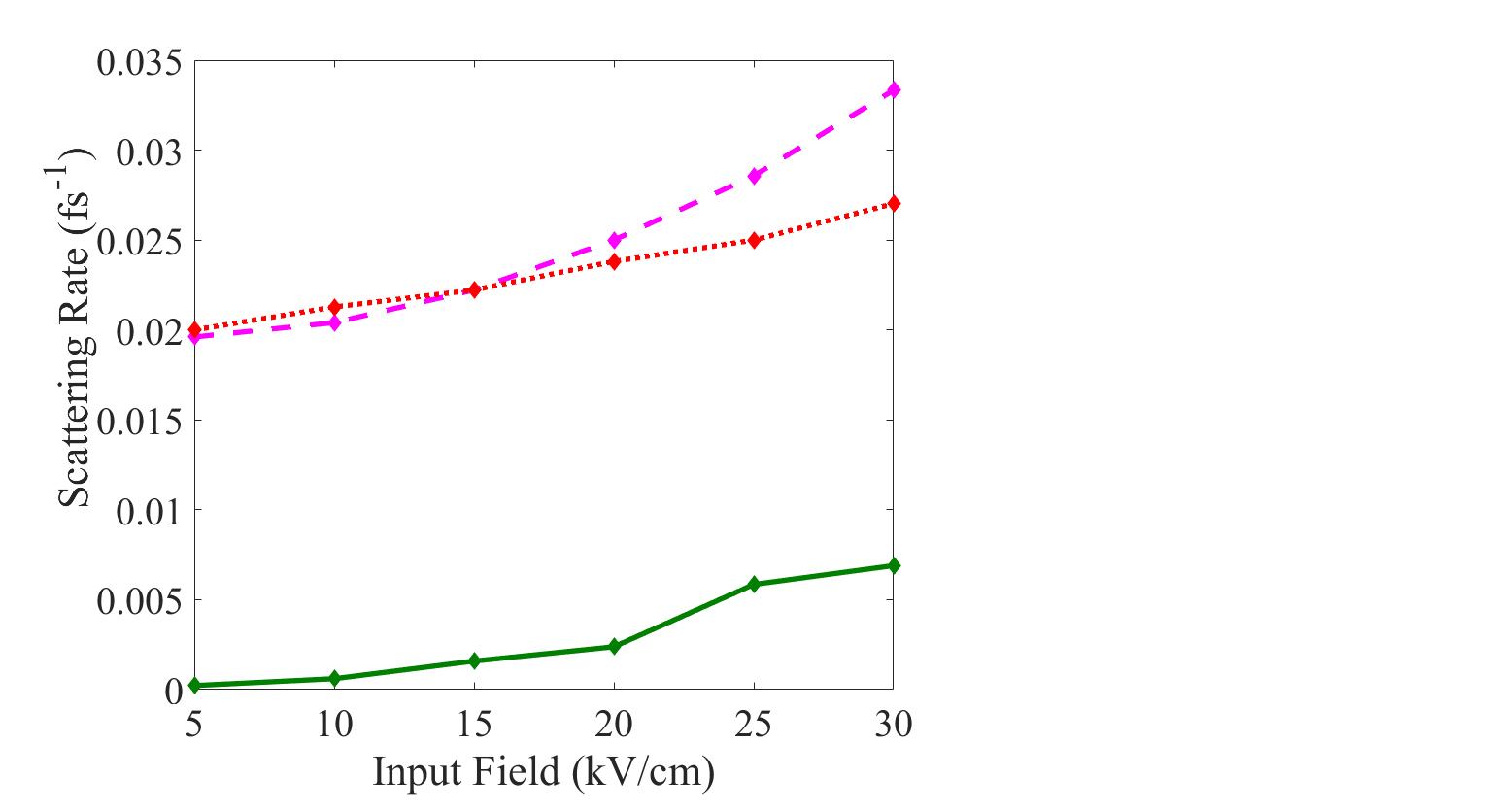} 
  \caption[Extracted effective scattering times]{Extracted effective scattering rate ($1/\tau$) as a function of input field for different scattering treatments; only phonon scattering (solid dark green), impurity scattering (dashed pink curve), and impurity and phonon scattering (dotted red curve).}
  \label{fig:ScattRate}
\end{figure}

 
 To aid in the understanding of the nonlinear transmission in the presence of microscopic scattering, let us consider the effects of the field amplitude on the microscopic scattering rates of the electrons. Increasing the input field amplitude pushes the electrons to higher energy states. From Eq. (\ref{eq_Gammain}), we see that this will lead to an increase in scattering due to neutral impurities and optical phonons. We now examine what happens for three different scattering scenarios. We use Eq. (\ref{eq_Etsimple}) with the nonlinear conductivity of Eq. (\ref{eq_SigNL}) to obtain the nonlinear transmission as a function of input field amplitude and scattering time. Then comparing these results to our simulation results, we extract an effective field-dependent scattering time for the different scattering scenarios and field amplitudes.  The resulting effective scattering times are plotted as a function of field amplitude in Fig. \ref{fig:ScattRate} \par

 Neutral impurity scattering increases when the field amplitude is increased because the number of states into which the electrons can scatter increases linearly with the electron energy and as electrons are driven to higher energy, there are more unoccupied states at the same energy into which they can scatter. When there is only neutral impurity scattering, the average scattering time decreases from $\tau =51$ fs at a field of 5 kV/cm to $\tau =30$ fs at a field of 30 kV/cm. \par  
 When there is only optical phonon scattering, the situation is somewhat different. The number of states into which the electrons can scatter due to interactions with phonons also increases as the electron energy increases. Thus, up to a field of about $20~\rm{kV/cm}$, there is a slow increase in the scattering rate and at $20~\rm{kV/cm}$ the scattering time reaches about 420 fs. However, as the field amplitude is increased further, some of the electrons have an energy that is at least an optical phonon energy above an empty state and phonon emission can suddenly occur. Thus, the scattering rate increases rapidly as the field is increased beyond 20 kV/cm and at the maximum field of $30~\rm{kV/cm}$, the effective scattering time is reduced to about $\tau =145$ fs. \par 
 The response when there is both neutral impurity and phonon scattering is somewhat more complicated. For low fields, neutral impurity scattering is dominant and the transmission is almost the same as what is found when there is only neutral impurity scattering. However, for fields above about $20~\rm{kV/cm}$, phonon emission processes kick in. This results in the average electron energy being lower than it would have been if only neutral impurity scattering were present. As a result, neutral impurity scattering rate decreases below what it would have been in the absence of phonon scattering, and so the transmission does not increase as much as it did when there was only impurity scattering present. At the highest field amplitude, the effective scattering time is about $\tau =37$ fs. \par 
\subsection{Third harmonic generation}\label{sec_Results_THG}
We now examine the dependence of the \textit{third harmonic response} on scattering, which is the central aim of this paper. As seen in Fig. \ref{fig:figure64}(a), even for a relatively low input field of $5~\rm{kV/cm}$, scattering has a strong effect on the generated third-harmonic field. Optical phonon scattering leads to the highest third-harmonic field and impurity scattering yields the lowest third-harmonic field of the microscopic models, independent of the input field amplitude. At low input field amplitudes, the third-harmonic field when there is both phonon and impurity scattering is slightly greater than the one with no scattering. However, as the input field amplitude increases, as shown in Figs. \ref{fig:figure64}(b) and \ref{fig:figure64}(c), the third harmonic is considerably lower when both impurity and phonon scattering are included. \par
\begin{figure}[h]
  \centering
  \includegraphics[ scale = 0.23]{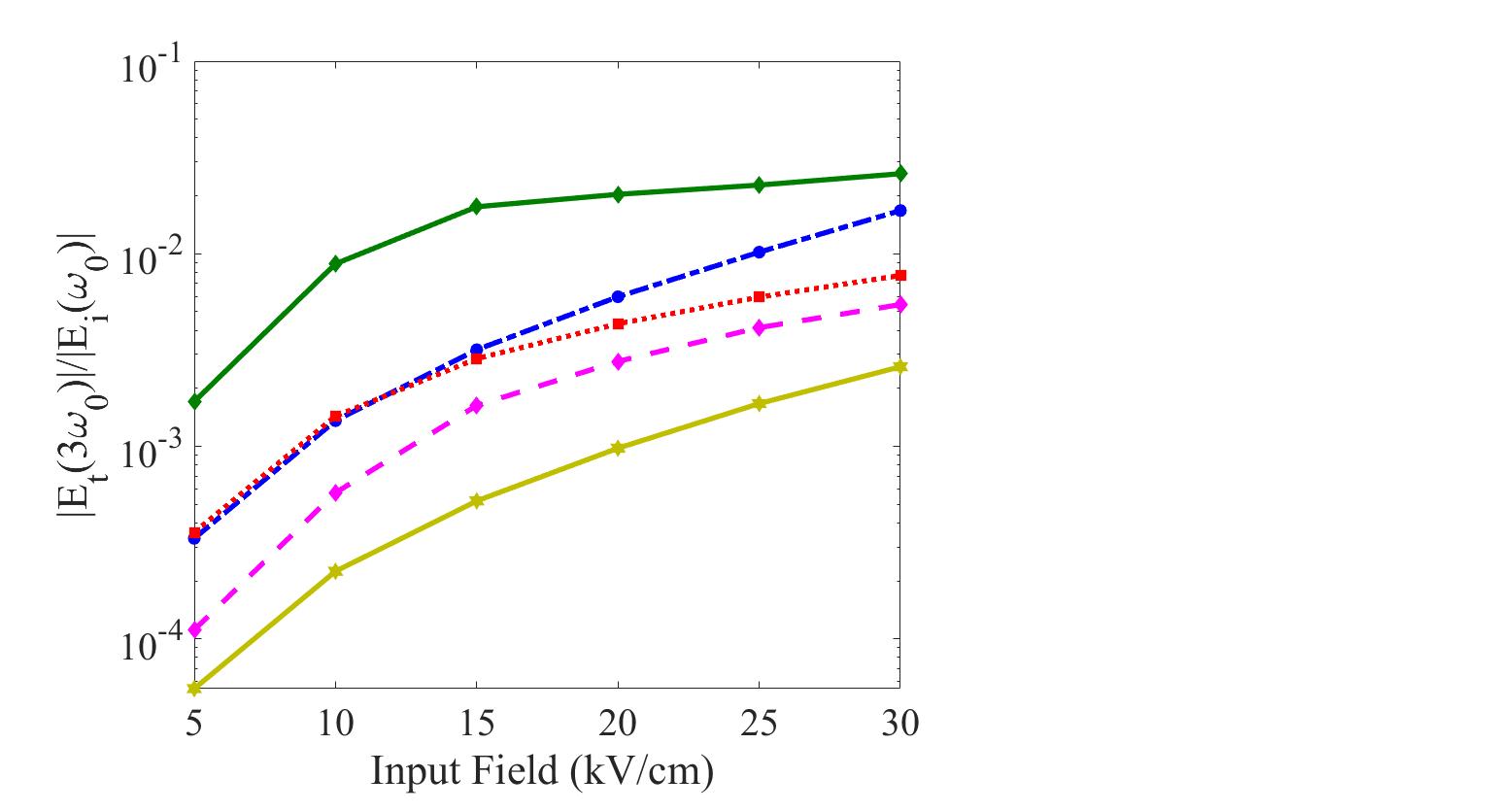} 
  \caption[Ratio of the transmitted third-harmonic field to the peak value of the incident field]{Ratio of the transmitted third-harmonic field to the peak value of the incident field as a function of input field for different scattering treatments; only phonon scattering (solid dark green), no scattering (dashed-dotted blue curve), impurity and phonon scattering (dotted red curve), impurity scattering (dashed pink curve), and semi-empirical model (solid light green curve).}
  \label{fig:figure68}
\end{figure}

 
To see the trends more clearly, in Fig. \ref{fig:figure68}, we plot ${\vert E_t( 3\omega_0)\vert} / { \vert E_i( \omega_0) \vert }$ as a function of input field amplitude for all the scattering scenarios. 
In all cases, we see an increase in the ratio of the generated third-harmonic field to the peak value of the incident field as we increase the input field amplitude. In a perturbative model in which the response is only calculated to third order in the field, the third-harmonic field amplitude should increase with the cube of the input field amplitude, which is not what we see in Fig. \ref{fig:figure68}. This difference arises from the field-dependence of the scattering times as well as higher-order nonlinearities that reduce the total field at the graphene, which results in a reduction in the third harmonic response \cite{naib2015}. Let us now consider each of the different scattering cases separately. \par  
When there is no scattering, the entire electron disc moves without distortion. Thus, it is in this case that the electrons are pushed to the highest energies. Therefore, one expects that perhaps this is when the largest nonlinear response would occur. We find that in this situation, the third harmonic is larger than in every other case, apart from when there is only optical phonon scattering. \par

The generated third harmonic field is the smallest when we employ the semi-empirical model of scattering. This is expected for two reasons. First, the process involves inelastic scattering, resulting in the carriers, on average, being scattered from higher energies back into the lower-energy thermal equilibrium distribution. This means that carriers are not driven as far from equilibrium, which in-turn reduces the generated third harmonic field. As a result, the third harmonic field amplitude is almost an order of magnitude lower than it was when there was no scattering. Now, when there is only neutral impurity scattering (which is elastic), carriers are scattered to different points with the same $\vert \mathbf{k} \vert$. We know from our examination of the transmission at the fundamental that in this case, as we increase the field amplitude, there is an increase in the scattering rate. Thus, we might expect that we would obtain similar results to those found for the semi-empirical model. However, we find that although there is a significant decrease in the third harmonic relative to that found when there was no scattering, the signal is much greater than that found for the semi-empirical model. This clearly shows that the effective scattering time itself is only part of the story and that the nature of the scattering is critical in determining its effect on third harmonic generation. The key here is that because the neutral impurity scattering is elastic, over time the Fermi disk gets larger and larger \cite{luke}, which means there are still many carriers that are driven to high energies by the field. \par
Now, let us consider the case of optical phonon scattering alone. Surprisingly, we see from Fig. \ref{fig:figure68} that this is the case in which the third harmonic field is the largest, particularly for the lower fields. This is a clear indication that the scattering process itself is resulting in a nonlinear component to the current that is even greater than that arising from the intrinsic band structure. \par
\begin{figure}
  \centering
  \includegraphics[scale=0.38]{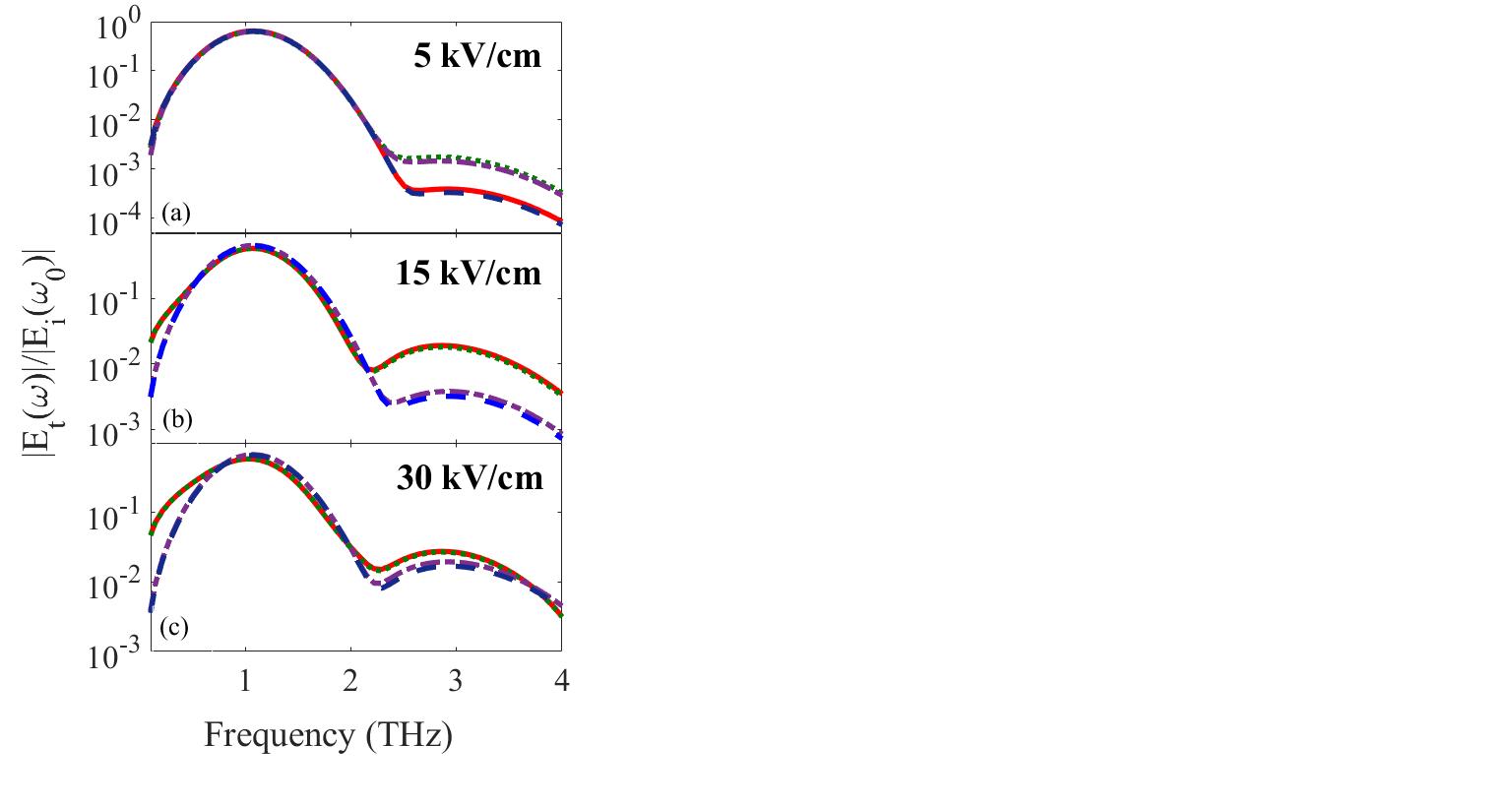} 
  \caption[Normalized transmitted field for different scattering treatments]{Normalized transmitted field for different scattering treatments: phonon emission and absorption (dotted green curve), only phonon absorption (dashed-dotted purple curve), only phonon emission (solid red curve), and no scattering (dashed blue curve) when the incident field peak amplitude is (a) $5~\rm{kV/cm}$, (b) $15~\rm{kV/cm}$, and (c) $30~\rm{kV/cm}$.}
  \label{fig:figure70}
\end{figure}

 
To better understand the third-harmonic response in the presence of only optical phonon scattering, in Fig. \ref{fig:figure70} we plot the normalized transmitted field for the case where there is only optical phonon scattering, but we examine the relative effects of phonon absorption and emission. Thus we plot the results under four different conditions: only phonon absorption, only phonon emission, both phonon emission and absorption, and no scattering. We see that at the low input field of $5~\rm{kV/cm}$, the effects of phonon emission are negligible and the result with only phonon emission is almost the same as when there is no scattering.  However, at the low field, phonon absorption significantly affects the third-harmonic field, even though the phonon populations at room temperature is very small ($n_{\Gamma} = 0.00051$ and $n_K = 0.0021$). The origin of the nonlinearity in this case seems to be the excitation of carriers to high-energy states, where the nonlinear relationship between the carrier crystal momentum and the velocity has the most pronounced effect. Although there are not many carriers with such high energies, because the intrinsic nonlinearity is so small at this low field amplitude, any small change in the current can have a large effect on the third harmonic. However, at the higher field amplitudes, the intrinsic nonlinearity is much stronger and it dominates over this phonon-induced contribution. As the input field amplitude increases, we see in Figs. \ref{fig:figure70}(b) and \ref{fig:figure70}(c) that phonon emission becomes the dominant scattering mechanism in determining the generated third-harmonic field. As we discussed earlier, when the incident field amplitude is large enough, carriers can be driven to high enough energy such that they can scatter into lower energy states by emitting an optical phonon. This will result in a clipping of the current similar to the intrinsic current clipping and the result is a larger third harmonic field than when there is no phonon scattering. \par 
The reason why THG is strong in the presence of optical photon scattering even though the linear response is suppressed can be understood by considering the carrier dynamics in k-space. The carriers that have the largest energy will be the ones that scatter to lower energy with the emission of an optical phonon.  These carriers will be the mostly close to the $k_x$-axis (\textit{i.e.}, close to $\theta =0,\pi$). From Eq. (\ref{eq_current3}), we see that the scattering of these carriers will reduce the linear conductivity. However, from the same equation, we see that this scattering does not significantly affect the third-order current, as the main contribution to that occurs for $\theta =\pm \pi /4, \pm 3\pi /4 $. If these carriers near $\theta =0$ are scattered into those regions in k-space, it can actually increase the nonlinear response. \par 

In the case that both neutral impurity and phonon scattering are included in the microscopic model, we see from Fig. \ref{fig:figure68} that for a low input field amplitude, the third-harmonic field is considerably larger than the case with only impurity scattering. This is because, as discussed above, the process of phonon absorption adds a significantly nonlinearity at these low fields that counter balances the reduction coming from impurity scattering. However, as the input field increases, the effect of the optical phonons is diminished. This is because at high fields, the elastic scattering due to neutral impurities is so much faster than the optical phonon scattering, many of the carriers that have been driven by the field to high energies are redistributed by neutral impurity scattering to k-states away from the leading edge of the Fermi disk before they are driven to energies high enough to emit an optical phonon.  \par
\begin{figure}
  \centering
  \includegraphics[scale = 0.25]{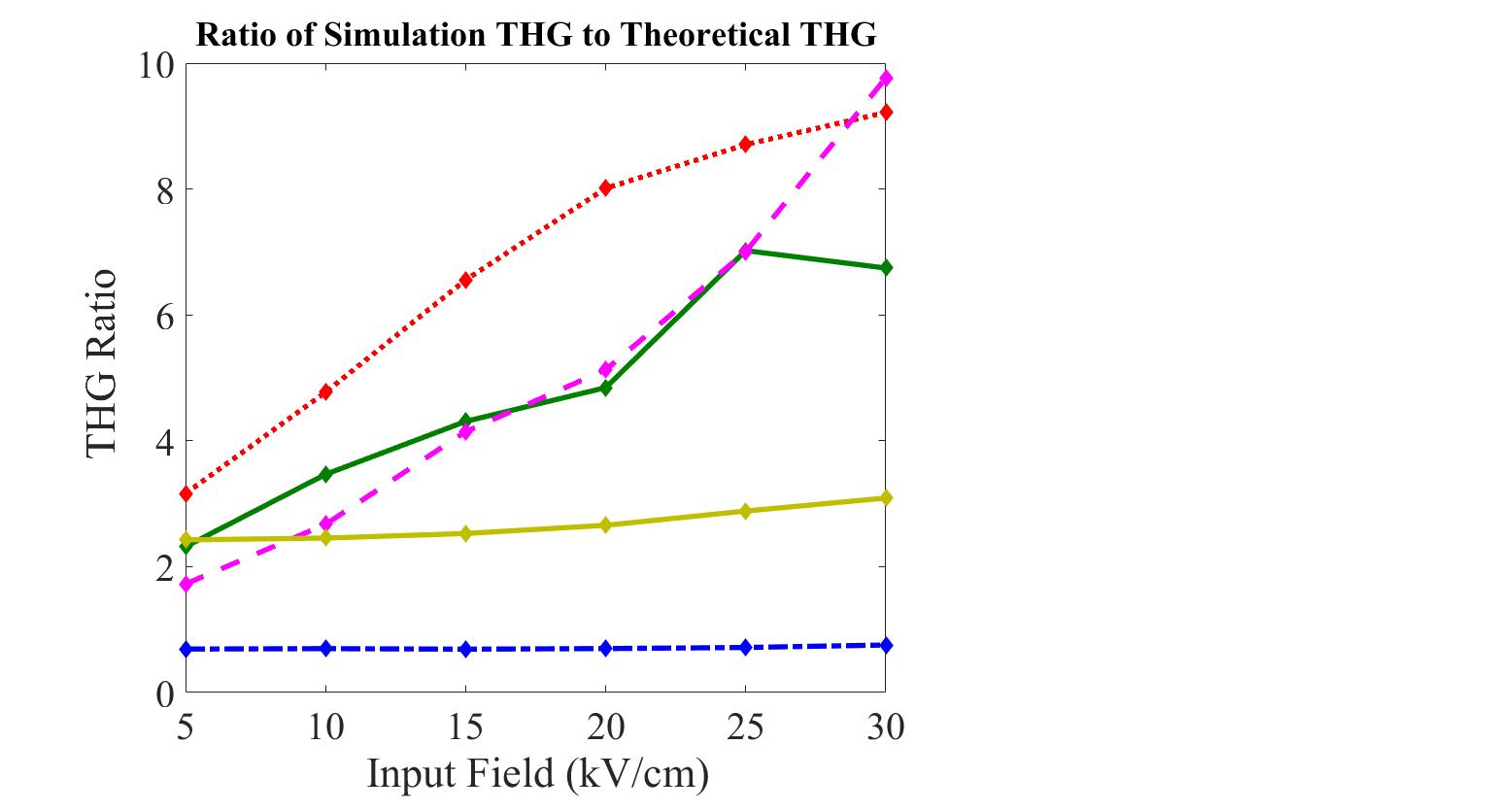} 
  \caption[Field dependence of the ratio of THG field amplitude from simulations that that from semiclassical theory]{Ratio of THG field amplitude from simulations to that obtained from semiclassical theory for different treatments of scattering; only phonon scattering (solid dark green), no scattering (dashed-dotted blue curve), impurity and phonon scattering (dotted red curve), impurity scattering (dashed pink curve), and semi-empirical model (solid light green curve).}
  \label{fig:Ratios}
\end{figure}
It is apparent from the results and the above discussion that the effects of scattering on THG cannot simply be captured by an effective scattering time and that the detailed carrier dynamics are crucial to understanding the results. To emphasize this point, we have calculated the third harmonic field found using the semiclassical model given in Sec. \ref{sec_SCtheory}, but using the effective field-dependent scattering time given in Fig. \ref{fig:ScattRate}. In Fig. \ref{fig:Ratios}, we plot the ratio of the third harmonic field calculated from the simulations to that calculated using this semiclassical model. The first thing to note is that the results using the two methods agree quite well when there is no scattering (as expected). When we consider the case where the scattering time is 52 fs for all fields (semi-empirical model), the simulation predicts a stronger third harmonic, but the dependence on field is very similar in both models. This is in contrast to the results when we include microscopic scattering. For these cases, although the results are similar at low field amplitudes, they are very different at higher fields. In particular, the results with neutral impurity scattering (with and without optical phonon scattering) differ by almost an order of magnitude at the highest field amplitude of 30 kV/cm. This is clear evidence that if one is to accurately calculate THG in graphene at THz fields, a microscopic model of the scattering is essential.

\section{Comparison to Experiment}\label{sec_Results_Expt}
In this section, we model the recent experiments of Hafez \emph{et al.} \cite{nature_hafez} using our microscopic model. In these experiments, they measured the THz harmonics in graphene in the time domain. The sample was monolayer graphene deposited on a silicon dioxide (SiO$_2$) substrate with a carrier density of $N_c = 2.1 \times 10^{12} ~\rm{cm}^{-2}$ (chemical potential of $\mu_c = 170~\rm{meV}$) at room temperature ($T = 300~\rm{K}$). The THz source used was the superconducting radio-frequency accelerator-based supperradiant THz source, TELBE \cite{free_laser}, with a pulse duration of about $14~\rm{ps}$ (or less), a central frequency of $f_0 = 0.68$~THz and a peak electric field amplitude of $12~\rm{kV/cm}$ to $85~\rm{kV/cm}$. In this section we compare the results of our simulation in the presences of neutral impurities and optical phonons to these experimental results. \par

\begin{figure}
  \centering
  \includegraphics[scale = 0.38]{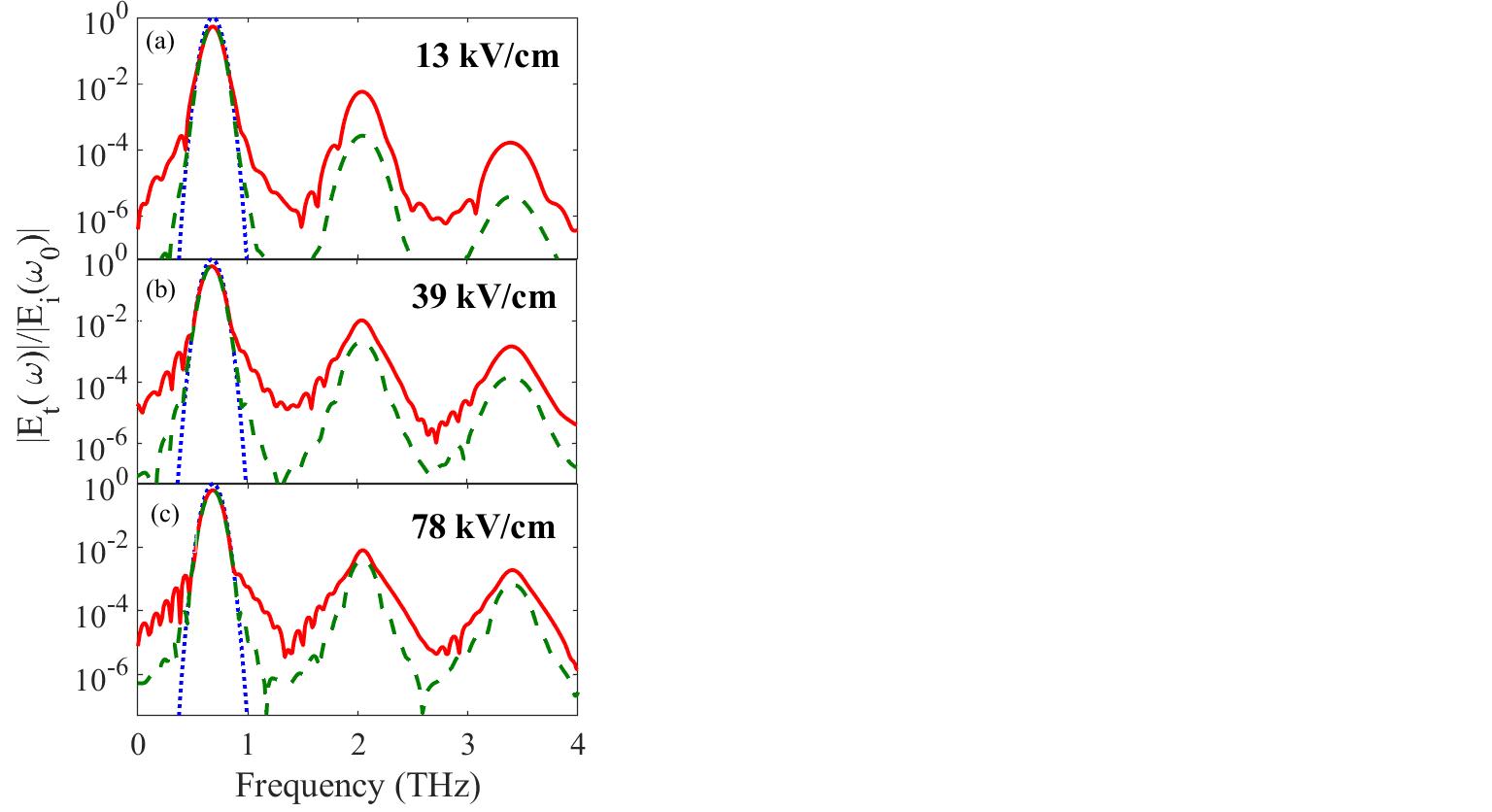} 
  \caption[The Fourier transformed incident field and transmitted field]{The Fourier transformed incident field and transmitted field as a function of frequency for input field amplitude of (a) $13~\rm{kV/cm}$, (b) $39~\rm{kV/cm}$, and (c) $78~\rm{kV/cm}$ using microscopic model (solid red curve) and semi-empirical model (dashed green curve). The dotted blue curve represents the incident filed. The fundamental frequency of the incident field is $0.68$~THz and both neutral impurity scattering and optical phonon scattering are included.  }
  \label{fig:exp1}
\end{figure}

In our simulations, we set $f_0 = 0.68$~THz, $T_{FWHM} = 7$~ps, $t_0 = 20$~ps. The pulse duration is somewhat shorter than that used in the experiment as the longer pulse simulations were too computationally-intensive: a typical calculation to produce one curve in Fig. \ref{fig:exp1}(c) takes $30$ processors $5$ days at the Centre for Advanced Computing. We have found, however, that our results do not change much when the pulse duration is increased. We take the index of refraction of the substrate to be $n=1.9$, which is the index of silicon dioxide at THz frequencies. We use a neutral impurity density of $n_{imp}  = 0.713~ \rm{cm}^{-2}$, chemical potential of $\mu_c = 170 ~\rm{meV}$, room temperature of $T = 300~\rm{K}$, and carrier density of $N_c = 2.1 \times 10^{12}~\rm{cm}^{-2}$, which yields a low-field scattering time of $47~\rm{fs}$, in agreement with the estimated scattering time in the experiment \cite{nature_hafez}. \\

In Fig. \ref{fig:exp1}, we plot the calculated transmitted field amplitude as a function of frequency field for input field amplitudes of $13~\rm{kV/cm}$, $39~\rm{kV/cm}$, and $78~\rm{kV/cm}$ using the microscopic model with both scattering mechanisms included, as well as the results using the semi-empirical model with $\tau = 47~\rm{fs}$. As can be seen, in addition to the large peaks at the third harmonic, we find peaks at the fifth harmonic. In table \ref{table:nonlin}, we present the extracted field amplitudes at the third harmonic from the experimental results and from our simulations for the three different input field amplitudes. As can be see, there is quite good agreement between the experimental results and the results from the semi-empirical model. There is also qualitative agreement for the input field of $78~\rm{kV/cm}$ between the experimental results and the results from our microscopic model. However, the results from the microscopic model seem to significantly overestimate the third-harmonic field for the two lower input fields.
Although the semi-empirical model gives better agreement with the experimental results, one should not read too much into this; this is clearly an overly-simplified model and the good agreement is likely fortuitous. \par 
\begin{table}[ht]
\caption{Third harmonic electric field amplitude for three different input field amplitudes found from experiment (column 2), using our full simulation with microscopic scattering (column 3) and with semi-empirical scattering (column 4).} 
\centering 
\setlength{\tabcolsep}{1.8pt}
\begin{tabular}{c c c c} 
\hline\hline 
Input Electric Field & Experiment  & Microscopic & Semi-Empirical \\ [0.5ex]
\hline 
13~kV/cm & 4.0~V/cm & 110~V/cm &  5.0~V/cm \\
39~kV/cm & 70~V/cm & 480~V/cm & 110~V/cm \\
78~kV/cm & 200~V/cm & 600~V/cm & 370~V/cm \\ [1ex] 
\hline 
\end{tabular}
\label{table:nonlin} 
\end{table}
There are a number of possible reasons why our results differ from the experimental results. First, there are some scattering mechanisms, such as electron-electron scattering, charged impurity scattering, acoustic phonon scattering that we have not included in our model, that may be important. The difference may also arise partly due to the particular estimates that we made for the phonon coupling constants and energies, on which there is no clear agreement in the literature \cite{Knorr}. Finally, uncertainties in the incident field and third-harmonic field in the experiment likely give some uncertainties in their results, which are not clearly identified in the paper. \par

\section{Conclusion}\label{sec_Conclusion}
In this paper, we have examined the effects of different scattering mechanisms on the nonlinear THz response of graphene. To make clear the effect of each scattering mechanism, we have investigated neutral impurity and optical phonon scattering individually and in combination. We have also compared these results to a model with constant scattering and a simple semiclassical model.\par
We have seen that the highest third-harmonic field is generated when there is only the optical phonon scattering, while neutral impurity scattering causes a decrease in the generation of the third-harmonic field. We have shown that even if one extracts a field-dependent effective scattering time from the nonlinear transmission at the fundamental frequency, this does not accurately capture the effect of the scattering on the generated third harmonic. This clearly shows that the microscopic details of scattering processes have a strong effect on third harmonic generation in graphene and must be taken into account if one is to obtain accurate predictions. Because the third harmonic signal is sensitive to the type of scattering, third harmonic generation in graphene at THz frequencies might be a very good way to characterize the type and strength of scattering in graphene.\par 
We have used our microscopic model to model the experiment recently done by Hafez \emph{et al.} \cite{nature_hafez} and obtain qualitative agreement for higher field amplitudes. To improve our model, in future work we plan to include other scattering mechanisms, such as electron-electron scattering and scattering from acoustic phonons.  \\\\
\textbf{ACKNOWLEDGEMENTS}\par
The authors would like to thank Luke Helt for his work in developing some of the theoretical and computational work used in this paper. We thank the Natural Sciences and Engineering Research
Council of Canada and Queen's University for financial support and the Centre for Advanced Computing for providing computational resources.

\end{document}